\documentclass[journal]{IEEEtran}

\IEEEoverridecommandlockouts
\usepackage{cite}
\usepackage{amsmath,amssymb,amsfonts}
\usepackage{graphicx}
\usepackage{textcomp}

\usepackage{scalerel}
\usepackage{tikz}
\usetikzlibrary{svg.path}

\definecolor{orcidlogocol}{HTML}{A6CE39}
\tikzset{
  orcidlogo/.pic={
    \fill[orcidlogocol] svg{M256,128c0,70.7-57.3,128-128,128C57.3,256,0,198.7,0,128C0,57.3,57.3,0,128,0C198.7,0,256,57.3,256,128z};
    \fill[white] svg{M86.3,186.2H70.9V79.1h15.4v48.4V186.2z}
                 svg{M108.9,79.1h41.6c39.6,0,57,28.3,57,53.6c0,27.5-21.5,53.6-56.8,53.6h-41.8V79.1z M124.3,172.4h24.5c34.9,0,42.9-26.5,42.9-39.7c0-21.5-13.7-39.7-43.7-39.7h-23.7V172.4z}
                 svg{M88.7,56.8c0,5.5-4.5,10.1-10.1,10.1c-5.6,0-10.1-4.6-10.1-10.1c0-5.6,4.5-10.1,10.1-10.1C84.2,46.7,88.7,51.3,88.7,56.8z};
  }
}

\newcommand\orcidicon[1]{\href{https://orcid.org/#1}{\mbox{\scalerel*{
\begin{tikzpicture}[yscale=-1,transform shape]
\pic{orcidlogo};
\end{tikzpicture}
}{|}}}}

\usepackage{kotex}

\usepackage[short,nodayofweek,level,24hr]{datetime}

\usepackage{algorithm}
\usepackage[noend]{algpseudocode}

\usepackage{pifont}

\usepackage{xcolor,color,soul}

\usepackage[hidelinks]{hyperref}

\usepackage{listings}

\usepackage{makecell}
\usepackage{tabu}

\begin{document}


\title{Thread Evolution Kit for Optimizing Thread Operations on CE/IoT Devices}

\author{

\IEEEauthorblockN{
Geunsik Lim$^{\textsuperscript{\orcidicon{0000-0003-1845-7132}}}$,  \IEEEmembership{Student Member, IEEE}, Donghyun Kang$^{\textsuperscript{\orcidicon{0000-0003-4362-9944}}}$, and Young Ik Eom$^{\textsuperscript{\orcidicon{0000-0001-6141-8054}}}$
}



\thanks{This work was partly supported by Institute of Information \& communications Technology Planning \& Evaluation (IITP) grant funded by the Korea government (MSIT) (IITP-2015-0-00284, (SW Starlab) Development of UX Platform Software for Supporting Concurrent Multi-users on Large Displays) and Basic Science Research Program through the National Research Foundation of Korea (NRF) funded by the Ministry of Science and ICT (NRF-2017R1A2B3004660).
(\textit{Corresponding author: Young Ik Eom}.)}
\thanks{G. Lim and Y. I. Eom are with the Department of Electrical and Computer Engineering, Sungkyunkwan University, 2066, Seobu-ro, Jangan-gu, Suwon, South Korea (e-mail: \{leemgs, yieom\}@skku.edu).}
\thanks{D. Kang is with Changwon National University, 20 Changwondaehak-ro Uichang-gu Changwon-si, Gyeongsangnam-do (51140), South Korea (e-mail: donghyun@changwon.ac.kr).}
}


\markboth{IEEE Transactions on Consumer Electronics, Vol. 1, No. 1, May 2020}{G. Lim \MakeLowercase{\text
{et al.}}: Thread Evolution Kit for Optimizing Thread Operations on IoT Devices}

\maketitle

\thispagestyle{empty}%


\begin{abstract}




Most modern operating systems have adopted the one-to-one thread model to support fast execution of threads in both multi-core and single-core systems. This thread model, which maps the kernel-space and user-space threads in a one-to-one manner, supports quick thread creation and termination in high-performance server environments. However, the performance of time-critical threads is degraded when multiple threads are being run in low-end CE devices with limited system resources. When a CE device runs many threads to support diverse application functionalities, low-level hardware specifications often lead to significant resource contention among the threads trying to obtain system resources. As a result, the operating system encounters challenges, such as excessive thread context switching overhead, execution delay of time-critical threads, and a lack of virtual memory for thread stacks. This paper proposes a state-of-the-art Thread Evolution Kit (TEK) that consists of three primary components: a CPU Mediator, Stack Tuner, and Enhanced Thread Identifier. From the experiment, we can see that the proposed scheme significantly improves user responsiveness (7x faster) under high CPU contention compared to the traditional thread model. Also, TEK solves the segmentation fault problem that frequently occurs when a CE application increases the number of threads during its execution.
 
\end{abstract}


\section{Introduction}\label{S_introduction}

\IEEEPARstart{A}s digital consumer electronics (CE) devices such as a smart refrigerator \cite{tce_iot_consumers} and smart television become common, it is important for traditional software layers to be optimized to mitigate the limitations of CE devices \cite{tce_senspnp, tce_ariot, tce_iot_pf2020}.
Nowadays, such devices are generally called IoT devices because they interoperate each other with internet facilities and sensor modules. Meanwhile, traditional operating systems of computing systems have adopted a model of one-to-one mapping between kernel-space and user-space threads because it allows opportunities for improving the scalability and performance of the system \cite{drepper2003native, wong2008fairness, Mueller93alibrary, boehm2005threads, nakashima2014massivethreads, qin2018arachne, barney2009posix, miller1999pk, engelschall2000portable, howell2013run, rieker2006transparent}. Unfortunately, this model does not fit for CE/IoT devices that have lower hardware specifications, because the model incurs some problems in that the threads running on CE/IoT devices often unintentionally spend a significant amount of time in taking the CPU resource and the frequency of context switch rapidly increases due to the limited system resources, degrading the performance of the system significantly. In addition, since CE/IoT devices usually have limited memory space, they may suffer from the segmentation fault \cite{segfault-clairvoyance} problem incurred by memory shortages as the number of threads increases and they remain running for a long time.

Some engineers have attempted to address the challenges of IoT environments such as smart homes by using better hardware specifications for CE/IoT devices \cite{tce_dev_sw, tce_ariot, tce_smart_log, vashisht2014study, icse-unified, dhotre2017analysis}. Unfortunately, this approach is inefficient and expensive because high-performance hardware requirement increases the manufacturing costs of CE/IoT devices. Other researchers and engineers have implemented {\em dual-version applications}: a generic version for normal computing systems and a light-weight version for CE/IoT systems \cite{icse-unified}. However, this approach also increases the cost of maintaining the applications because both versions of the software code must be modified when a software update is made. Meanwhile, in traditional systems, there is no concept of thread priority, and thus, it is difficult to identify time-critical threads that require quick responsiveness. As a result, when many active threads are running at the same time, all the threads, including those that are time-critical, unprejudicedly compete for system resources. This leads to performance collapse along with dropping user responsiveness.

This paper proposes a \textit{Thread Evolution Kit (TEK)} that adopts the modern one-to-one thread model for CE/IoT devices while leveraging its benefits in the user space. TEK is composed of three primary components: a CPU Mediator, Stack Tuner, and Enhanced Thread Identifier. First, we designed the CPU Mediator to help developers set the priority of time-critical threads. TEK is also implemented with a priority-based scheduler that isolates high-priority threads (i.e., time-critical threads) from normal ones. The goal of the Stack Tuner is to determine how much memory space should be allocated for thread stacks. Also, it is used to avoid the problem of {\em coarse-grained stack management} in which existing operating systems give each thread more stack memory than is actually used by the thread. To implement the Stack Tuner, this paper revisited the existing stack management mechanism and designed a scheme that could automatically assign an appropriate stack size by obtaining the thread’s actual stack usage. With the proposed scheme, software developers for CE/IoT devices can easily use TEK to prevent virtual memory shortages. Meanwhile, to employ TEK, software developers must specify information on the threads using the special APIs provided by TEK. In order to correctly handle each thread, the Enhanced Thread Identifier inspects this information whenever the program codes are compiled. For evaluation, TEK was implemented on an CE/IoT development board and compared with the conventional system. Surprisingly, the results show that the response time of time-critical threads was accelerated by up to 7x, and the amount of memory space was saved by up to 3.4x, compared with the conventional software platform. 

The remainder of this paper is organized as follows. Section \ref{S_background} describes the strong and weak points of each thread model. Section \ref{S_observation} discusses the observation results of conventional thread operations on CE/IoT devices. Section \ref{S_design} presents the design and implementation of the proposed schemes, and Section \ref{S_evaluation} shows the evaluation results. Related work is described in Section \ref{S_related_work}, and finally, Section \ref{S_conclusion} concludes the paper.


\section{Background}\label{S_background}
This section compares the strong and weak points of the three major thread models: many-to-one, many-to-many, and one-to-one, as shown in Fig. \ref{fig:thread-models-compare}. Then, it addresses how modern operating systems such as Linux have evolved the thread mapping model.

\subsection{Comparison of Thread Models}\label{SS_comp_thread_models}

\begin{figure}
\centering
\includegraphics[width=0.99\columnwidth,height=1.7in]{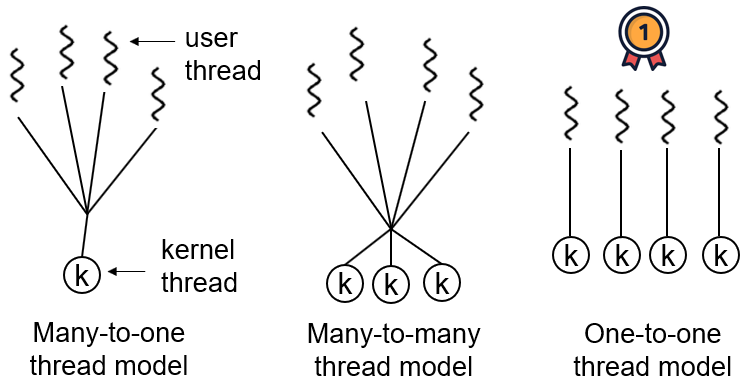}
\caption{Three types of thread models. Popular operating systems \cite{2016rethink, drepper2003native, 2013win32, 2010evolution} adopt the one-to-one thread model as their major thread model.}
\label{fig:thread-models-compare}
\end{figure}

The model for thread mapping between the user space and kernel space has a great influence on the behavior of the threads from their creation to termination. Fig. \ref{fig:thread-models-compare} shows the different operation flows of the common thread models and how each works. This section addresses the features, strengths, and weaknesses of each thread model.

\textit{Many-to-one thread model:} Many operating systems have commonly used this model as it allows for simple implementation and portability. In addition, this model provides for much quicker context switching compared with other models because all threads are managed in the user space. Unfortunately, when this model is applied, a thread can block the flows of other threads for a long time. For example, if one thread triggers a system call that leads to resource contention, other threads of the same application must also wait until the thread gets the resource. Moreover, this model cannot fully take advantage of the benefits of a multi-processor architecture because it handles only one application with a single processor, even though the application is composed of multiple threads \cite{engelschall2000portable, miller1999pk}.

\textit{Many-to-many thread model:} This model allows for parallel mapping between the user-space and kernel-space threads. In this model, most context switching operations occur in the user space, thus, this model shares kernel objects among the user-space threads. However, this model requires more effort to implement than the many-to-one model due to its complicated structure. Also, now that priority is assigned to the threads through the thread manager in the user space, the threads cannot directly access the kernel objects to obtain CPU resources \cite{franke2002fuss, brown2007c++}.

\textit{One-to-one thread model:} Most modern operating systems employ the one-to-one thread model. This model manages threads directly in the kernel space so that kernel-level scheduling is possible for each thread. In multi-processor systems, multiple threads can run on different CPUs simultaneously, and there may be no delay, even during blocks of system services. When one thread calls a blocking system call, the other threads do not need to be blocked. However, since this model performs context switching in the kernel space, the context switching operation is relatively slow compared to the many-to-one model. Also, with each thread taking up kernel resources, if an application creates many threads, the performance of the threads depends heavily on the hardware specifications \cite{drepper2003native, harrow2000runtime, pouget2010user}.

It is now common for an application to be developed by dozens of developers due to the increased size and complexity of applications, and so, the number of threads in modern CE/IoT devices has increased from tens to hundreds. Meanwhile, when the number of user-space threads in one application increases to hundreds, it is difficult for the application developer to know the roles of threads created by other developers because the existing thread models do not provide developers with a mechanism to monitor thread functionalities \cite{thread-improving, harrow2000runtime, pouget2010user}. As a result, it becomes increasingly difficult to control and optimize their behavior.

\subsection{Evolution of Linux Thread Model}

In recent decades, developers implemented a multi-threaded application with the many-to-one thread model. The many-to-one thread model made thread implementation easy by providing application developers with intuitive user-space thread operations as discussed in Section \ref{SS_comp_thread_models}. However, user-level threads based on the many-to-one thread model could not utilize multi-processors directly without kernel-level support from the operating system. As a result, with the advent of multi-processor systems, operating systems added kernel-level support for threads \cite{engelschall2000portable, drepper2003native}.

The Linux kernel introduced Fast User-Level muTEX (FUTEX) \cite{franke2002fuss} to support light and fast synchronization. This can be used when the threads access shared resources in the multiprocessor environment. The system calls of the FUTEX facility provide application developers with a fast user-level synchronization mechanism \cite{gidenstam2007lfthreads, drepper2005futexes, holman2006locking}. The one-to-one thread model of Linux supports real-time FUTEX behavior for user-space threads. Then, the Linux kernel adopts LinuxThread, which uses a \textit{clone} system call to enable user-space real-time thread operations \cite{drepper2003native, franke2002fuss}. Since the \textit{clone} system call of the one-to-one thread model provides a mechanism that tracks threads to speed up thread creation, Linux dramatically improves the scalability and performance of thread execution. Also, now that the one-to-one thread model eliminates the user-level thread manager for fast thread operation, it both simplifies the operation flow of the threads and significantly accelerates the execution speed of thread termination. As a result, applications no longer depend on a thread manager that may cause context switching and performance degradation \cite{drepper2003native, qin2018arachne, adya2002cooperative}.

The one-to-one threaded programs can run even faster since Linux introduced the O(1) scheduler, which consists of run queues and bitmap priority arrays. This improves scalability without adding a performance penalty in multi-processor environments \cite{wong2008fairness, thibault2005flexible}. As a result, when the O(1) scheduler allowed threads to run quickly in high-performance multi-core server environments, modern operating systems adopted the one-to-one thread model as its standard thread model.

Finally, the one-to-one thread model safely solves existing POSIX compliance problems because it performs signal handling of the threads in kernel space. Moreover, because Linux maps the user-space thread to the Light-Weight Process (LWP) in kernel space, it completely links the system resource usage of the thread to that of the LWP of the Linux kernel. On Linux, the LWP refers to processes sharing the same memory address space and other resources in the kernel space. Therefore, the thread library can correctly monitor the thread behaviors of an application using the pseudo filesystems (e.g., \textit{proc} and \textit{sysfs}) \cite{procfs-sysfs} in Linux.

Even though the Next-Generation POSIX Threads (NGPT) \cite{franke2002fuss} proposed a many-to-many model in which many user threads are mapped onto many kernel threads, unlike the existing one-to-one thread model of the Linux kernel, it cannot solve all the problems of the user-space library threads. The main reason that the traditional LinuxThread has been used as the dominant thread library for so long time is that the kernel-level threads of the operating system solve fault handling and thread performance problems.

\section{Observation}\label{S_observation}

This section discusses the observation that modern CE/IoT devices bring new challenges to the existing one-to-one thread model.

\subsection{CPU Contention Among the Threads on Low-End CE/IoT devices}

The latest Linux kernel supports two types of CPU schedulers: the O(1) and the Completely Fair Scheduler (CFS) \cite{dhotre2017analysis}. The O(1) scheduler provides CPU scheduling of real-time threads with a First-In-First-Out (FIFO) or Round-Robin (RR) scheduling policy, and controls each thread according to its fixed priority. On the other hand, the CFS scheduler supports fair scheduling of non-real-time threads with a NORMAL scheduling policy, controlling each thread according to its dynamic priority \cite{wong2008towards}. The fixed priority does not change while scheduling threads, while strictly maintaining the scheduling sequence. On the contrary, the dynamic priority changes from time to time while scheduling threads because CFS dynamically recalculates the weight values of the threads in the system. Typically, user-space applications create new threads that are controlled by the NORMAL scheduling policy, and they are scheduled in a time-sharing manner. In detail, the NORMAL scheduling policy manages CPU resources using the virtual run-time \cite{kim2018fair} with the red-black tree which is usually used for efficient self-balancing binary search \cite{redblack-tree} to ensure that all threads use CPU resources fairly \cite{wong2008towards}. The CFS scheduler adopts the notion of virtual run-time, which should be equal for all tasks, where the virtual run-time normalizes the value of the real run-time of a given thread with its \textit{nice} value (user-level thread priority). At this time, the CFS scheduler employs a red-black tree to support efficient self-balancing binary search algorithm. The red-black tree, in combination with the virtual run-time, ensures that higher priority tasks gain access to CPU resources more frequently without starving the lower priority tasks.

However, the CFS scheduler does not consider low-end CE/IoT devices, especially when the CE/IoT device has to execute time-critical threads with more favor under high CPU contention. The \textit{nice} value of -20 to 19 entered by the application developer is based on the 40 weight values defined by CFS as an array. For example, if a developer creates four threads with \textit{nice} values of 1, 2, 3, and 4, the CPU usage becomes 26.5\%, 25.5\%, 24.5\%, and 23.5\%, respectively, because the operating system applies equitable weight values to the threads. In other words, the CFS scheduler focuses on fair resource distribution among the threads in the system. Therefore, if the threads frequently compete to obtain available CPU resources, the existing scheduler decreases the response time of the time-critical threads in which user responsiveness is important.

\begin{figure}
\centering
\includegraphics[width=0.99\columnwidth,height=1.5in]{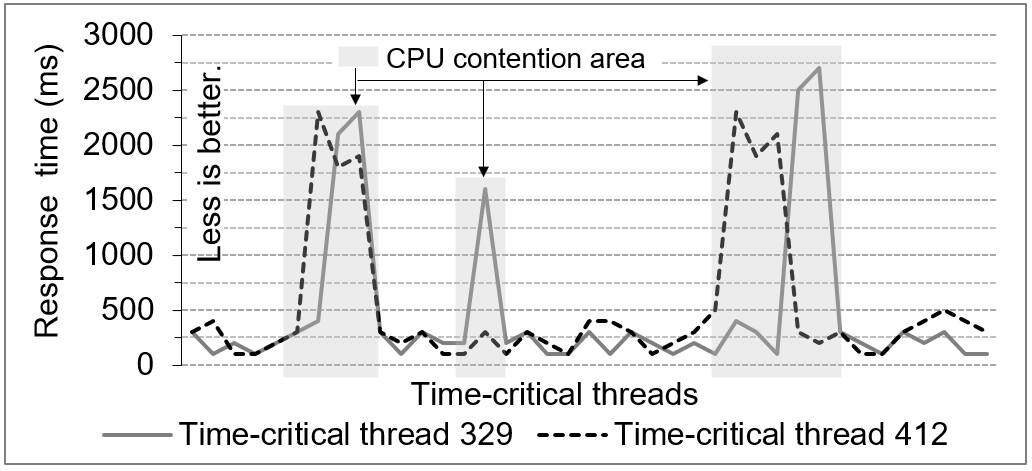}
\caption{The response time of two time-critical threads in an urgent group during CPU contention. The gray rectangle represents a region where threads compete for available CPU resources. Time-critical thread 412 probes the current temperature from a thermal peripheral device, while time-critical thread 329 displays the temperature.}
\label{fig:observ-time-critical-tid-slow}
\end{figure}

Fig. \ref{fig:observ-time-critical-tid-slow} shows the delay in processing time-critical threads that measure the current temperature during high CPU (Embedded CPU quad-core 1.2GHz) contention in a real CE/IoT system environment such as a refrigerator. In the experiment, the time-critical threads and the background service threads in the same thread group were intensely competing with one another for CPU resources. As a result, the system frequently reproduced unpredictable processing delays for the time-critical threads. Even though high-end hardware (e.g., X86 CPU) can minimize the frequency of CPU contention compared to low-end hardware (e.g., Embedded CPU), most CE/IoT devices require an energy-efficient CPU for low power consumption and smaller die size. Therefore, it is crucial to optimize the processing speed of time-critical threads under high CPU contention in low-end devices. In Section \ref{S_design}, this paper describes the design and implementation of the technique proposed to solve this problem in detail.


\subsection{Segmentation Fault of New Threads on CE/IoT devices}\label{SS_observ_mem}

In general, a segmentation fault occurs when a running thread accesses an illegal memory location or tries to write onto a read-only memory location \cite{segfault-clairvoyance}. Surprisingly, it is observed that there is another case when segmentation faults are generated while running an application. When an application requests to create a new thread, the operating system builds a new stack on the virtual memory space and then clones the contexts of the parent process, such as code and data, to that of the new thread. However, whenever a thread is created, the existing operating system gives each thread more space in stack memory than the amount of space the thread actually uses (e.g., in Linux, a stack space of 8 MB or 12 MB is automatically assigned for each thread). Therefore, this coarse-grained stack management problem, which induces a lack of system stack space, may be accelerated over time. Unfortunately, in 32-bit architecture, it is not easy to solve this problem because an application can use a total of 3 GB for the user-space area. For example, if an application simultaneously runs 300 threads with 8 MB stack for each thread, the existing operating system may incur a segmentation fault when creating a new thread. Of course, the frequency of segmentation faults depends somewhat on how the software platform handles the virtual memory area.

\begin{figure}
\centering
\includegraphics[width=0.99\columnwidth,height=2.2in]{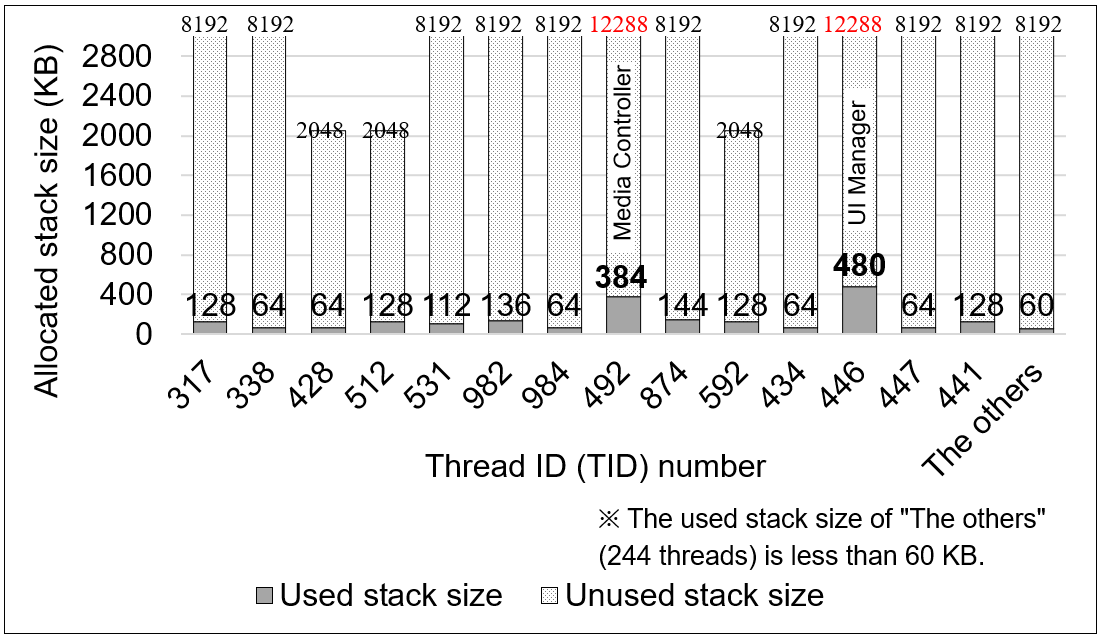}
\caption{The actual stack usage of 258 user-space threads on a CE/IoT device.}
\label{fig:observ-thread-stack-usage}
\end{figure}

Fig. \ref{fig:observ-thread-stack-usage} shows the actual stack usage of 258 user-space threads that are run on a CE/IoT device (1 GB RAM LPDDR2 900MHz, 4 GB virtual memory with the memory management unit) with coarse-grained stack management. In Linux, a new thread requires a minimum stack size of 16 KB to establish a data structure of the user-level thread. As shown in Fig. \ref{fig:observ-thread-stack-usage}, most of the threads allocate much higher stack size than they use in reality. From the analysis, although all threads run tasks with the default stack size (8 MB) of the system, more stack space is allocated for the UI Manager (TID 446) and Media Controller (TID 492). On the contrary, less stack space is allocated for the threads of TID 428, 512, and 592. The reason for this is that the developers employ the \textit{pthread\_attr\_setstacksize} API \cite{barney2009posix} in order to directly manipulate the stack space of the threads.

These observations give further motivation to propose TEK because it is believed that it is possible to resolve the resource management problems incurred due to excessive resource contention while running the one-to-one mapping model between kernel-space and user-space threads.

\section{TEK: Design and Implementation}\label{S_design}

\begin{figure}[ht]
\centering
\includegraphics[width=0.99\columnwidth,height=2.2in]{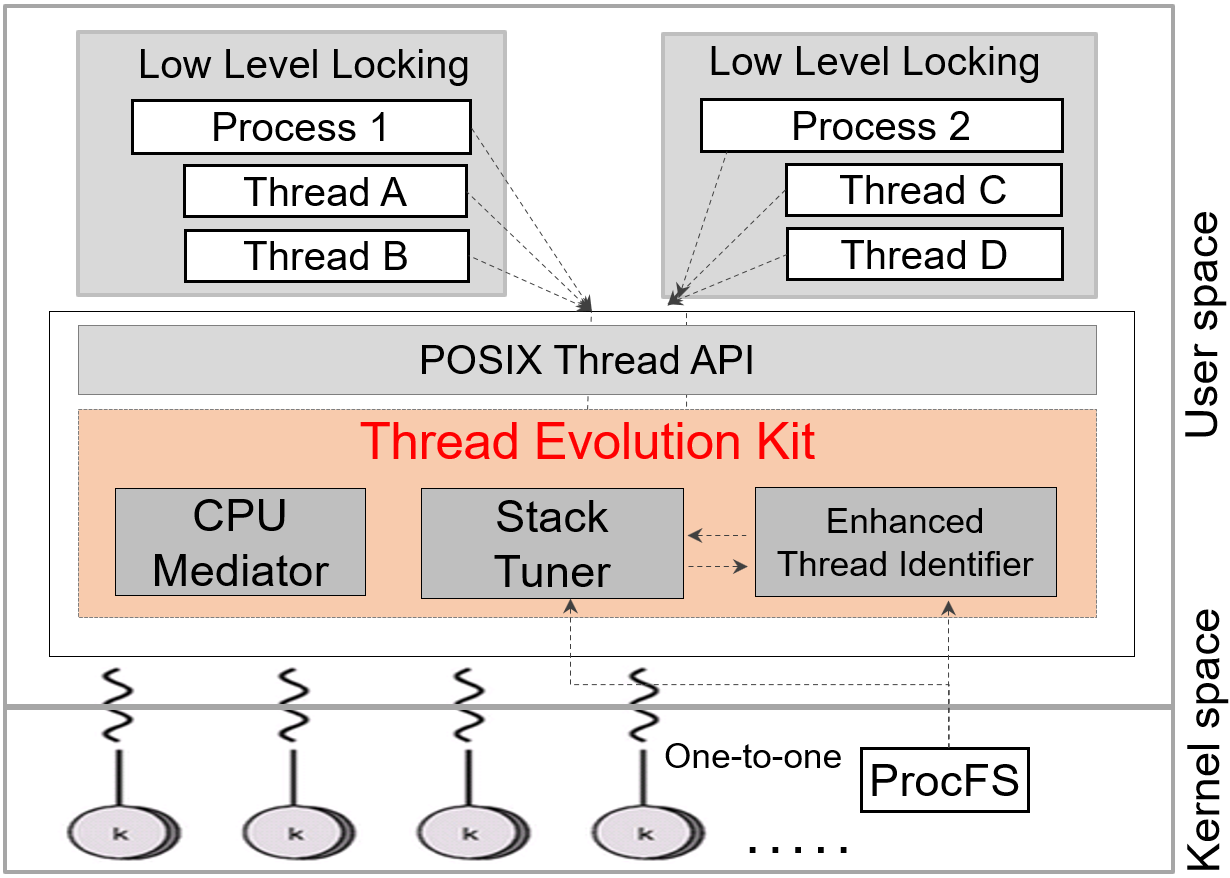}
\caption{Overall architecture and operation flow of TEK.}
\label{fig:system-architecture}
\end{figure}

\begin{table}
\caption{The POSIX APIs for CPU scheduling of processes and threads}
\begin{tabular}{llll}
\hline
\text{API name}        & \text{Arguments}          & \text{Who} & \text{Type}   \\ \hline
nice          & \ding{182}inc                & Process & Syscall \\ \hline
setpriority   & \ding{183}which, \ding{184}who, \ding{185}prio   & Process & Syscall \\ \hline
\makecell{pthread\_set\\schedparam} & \ding{186}thread, \ding{187}policy, \ding{188}priority & Thread   & Libcall \\ \hline
\end{tabular}
\\
\ding{182} inc: a nice value for the calling process.\\
\ding{183} which: PRIO\_PROCESS, PRIO\_PGRP, or PRIO\_USER.\\
\ding{184} who: a process group or real user ID of the calling process.\\
\ding{185} prio: a value in the range -20 to 19. The default priority is 0.\\
\ding{186} thread: a thread ID.\\
\ding{187} policy: a scheduling policy (e.g., SCHED\_TEK for CPU Mediator).\\
\ding{188} priority: a scheduling priority (e.g., a nice value for SCHED\_TEK).\\

\label{tab:posix-api-for-sched}
\end{table}

\begin{figure*}
\centering
\includegraphics[width=1.6\columnwidth,height=2.2in]{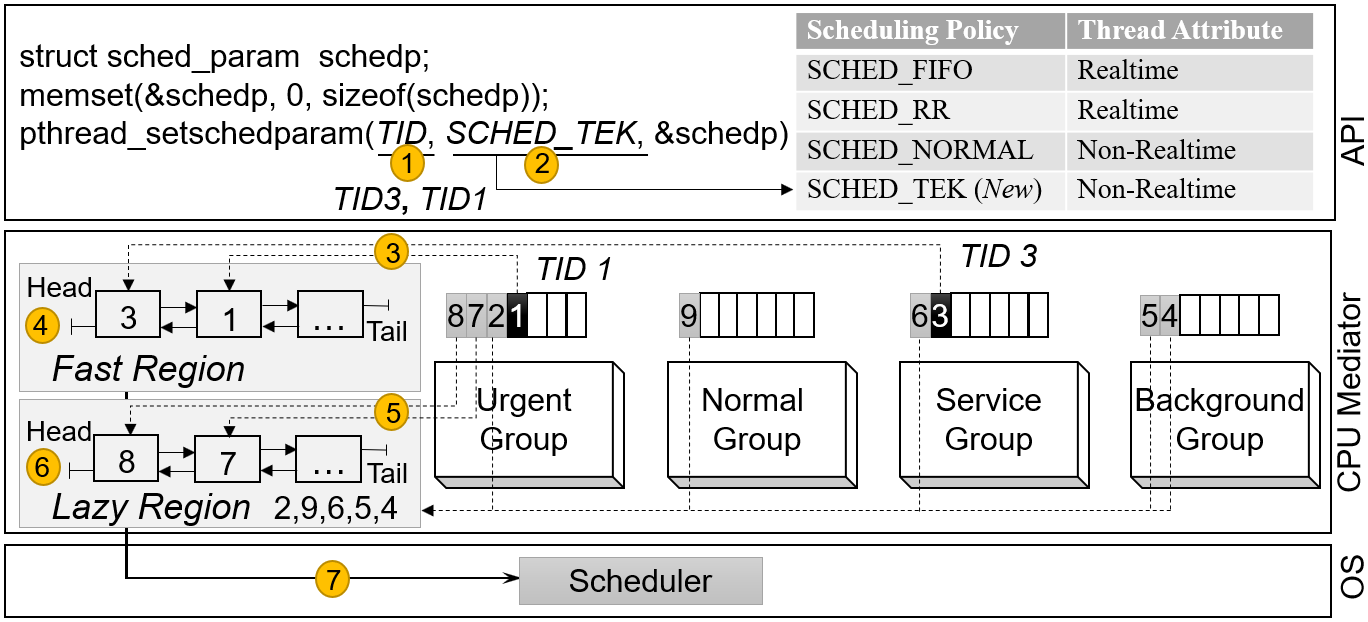}
\caption{Thread programming model with SCHED\_TEK of CPU Mediator for accelerating the response time of time-critical threads.}
\label{fig:time-critical-thread-controlling}
\end{figure*}

This section introduces the  \textit{Thread Evolution Kit (TEK)}, designed in the same spirit as the traditional one-to-one thread model to handle application threads in the user space. However, to develop applications in CE/IoT environments without re-design of the existing one-to-one thread model, TEK optimizes the previous thread model with three key components:

\begin{itemize}
\item \textit{CPU Mediator} (Section~\ref{SS_cpu_mediator}): This component supports fine-grained thread management in which each thread is handled based on the priority given by the application developers.
\item \textit{Stack Tuner} (Section~\ref{SS_stack_tuner}): The goal of this component is to optimally allocate stack memory in the virtual address space whenever an application creates a new thread.
\item \textit{Enhanced Thread Identifier and New APIs} (Section~\ref{SS_thread_mm}): This component is responsible for handling the hundreds of threads running on a CE/IoT device and provides new APIs to designate a thread as time-critical or non-time-critical.
\end{itemize}

Fig. \ref{fig:system-architecture} shows how applications are managed with TEK. TEK provides application developers with POSIX-compatible thread APIs (e.g., \textit{pthread\_setschedparam}) that support optimization of resource management in both low-end CE/IoT devices and existing high-end server systems (i.e., in TEK, the improved \textit{pthread\_setschedparam} API is used to run a unified application that is compatible with both low-end and high-end devices). Now, this paper discusses the key components of TEK in detail.

\subsection{CPU Mediator}\label{SS_cpu_mediator}

The existing software layer for CE/IoT devices was designed to handle threads with a group scheduling policy (i.e., coarse-grained thread management). Because modern applications create more and more threads, this technique can effectively control CPU resources by grouping the threads of each application. However, this coarse-grained thread management technique may be harmful in modern CE/IoT environments in which threads may require fast responsiveness because it is not easy to predict which thread will be run next.

The CPU Mediator is designed to support fast and predictable thread execution. In particular, the CPU Mediator classifies all threads running on a CE/IoT device into two categories according to their priority: time-critical and non-time-critical. The scheduling priority (\ding{188} in Table~\ref{tab:posix-api-for-sched}) of each thread is set by calling the APIs supported by TEK as described in Table \ref{tab:posix-api-for-sched} and would not be adjusted until the thread terminates. For time-critical threads, this paper further implements a new scheduler policy, called SCHED\_TEK (\ding{187} in Table~\ref{tab:posix-api-for-sched}), that offers more chances to obtain CPU resources by delaying non-time-critical threads.

This paper considers an example scenario in which time-critical threads are processed. When an application runs time-critical threads to guarantee fast response time, the CPU Mediator changes the policy of the scheduler to SCHED\_TEK by calling the \textit{pthread\_setschedparam} API in user space. Fig. \ref{fig:time-critical-thread-controlling} shows a logical thread migration flow of the CPU Mediator along with the SCHED\_TEK policy. The SCHED\_TEK policy starts to control the CPU resources according to the following two steps. First, the CPU Mediator looks up the time-critical threads in the group where the user-space thread lays based on its kernel-space thread ID\footnote{The kernel-space thread ID is acquired by using the modified {\em gettid()} system call where this paper replaces FUTEX with the Read-Copy-Update (RCU) mechanism because it is more suitable for read-intensive operations.}, and then logically migrates them to the {\em Fast Region}, where the probability of obtaining CPU resources is relatively high, as shown in Fig. \ref{fig:time-critical-thread-controlling}. Second, the CPU Mediator dynamically drops the priority of each non-time-critical thread running on the CE/IoT device to yield CPU resources to the time-critical threads. Then, it logically migrates all non-time-critical threads to the {\em Lazy Region}, where the execution of the threads will be delayed until the {\em Fast Region} becomes empty. The purpose of the {\em Fast Region} is to accelerate the processing speed of time-critical threads, while that of the {\em Lazy Region} is to delay the other threads. These regions link or unlink the threads of the existing groups with a doubly-linked list. After the time-critical threads in the {\em Fast Region} are terminated, the CPU Mediator unlinks the threads belonging to the {\em Lazy Region} and puts them into their original groups, instantly restoring their scheduling policy.

\begin{figure}[ht]
\centering
\includegraphics[width=0.99\columnwidth,height=2.7in]{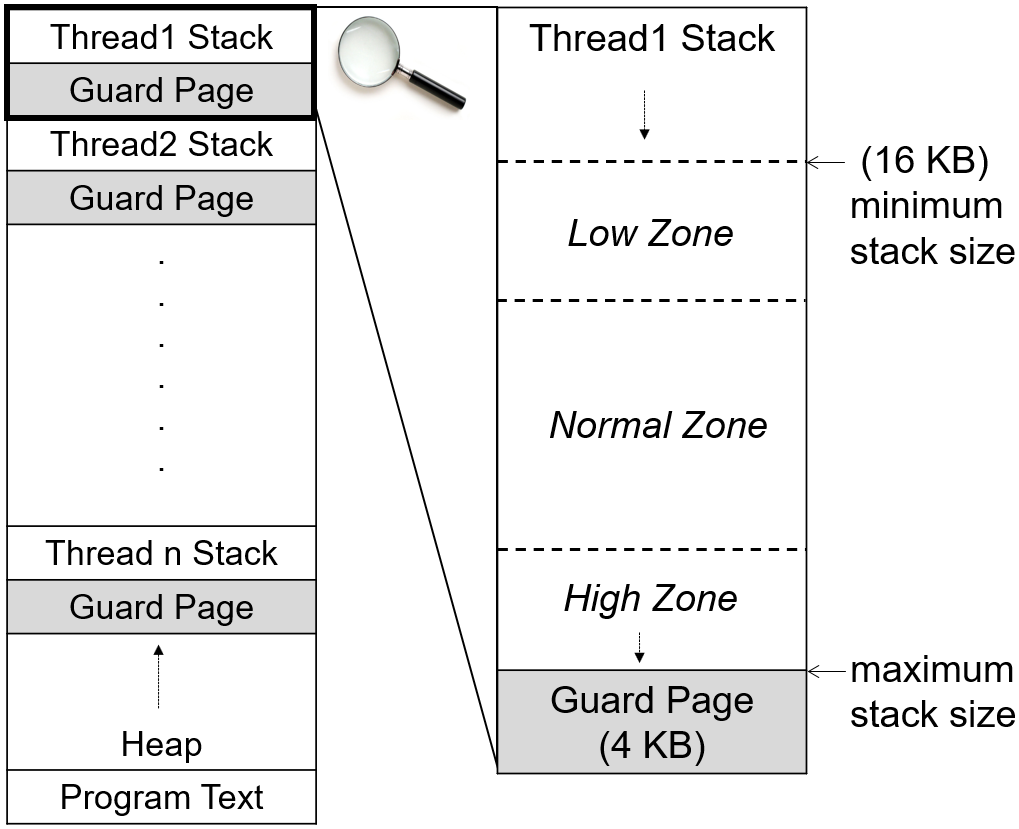}
\caption{The stack management structure of the Stack Tuner used to avoid a shortage of virtual memory.}
\label{fig:memory-tuner}
\end{figure}

\subsection{Stack Tuner}\label{SS_stack_tuner}

As the number of threads increases, segmentation faults may occur frequently due to the lack of system stack space in the virtual memory. Traditional operating systems always allocate stack space as requested by the thread\footnote{{\em The pthread\_attr\_setstacksize()} API is used to explicitly allocate the stack memory space.},  regardless of how much stack space is actually used by the thread at runtime; this situation is very similar to the {\em internal fragmentation issue} in physical memory. For example, if a thread running for an application only uses 1 MB of stack space after being allocated 100 MB of stack space, a significant amount of virtual memory (i.e., 99\%) is wasted. Therefore, the coarse-grained stack management technique mentioned in Section \ref{SS_observ_mem} may accelerate the lack of the system stack space over time.

To address the lack of system stack space, this paper designed the Stack Tuner to monitor the stack space during the lifetime of each thread. In order to measure the stack usage of each thread, the Stack Tuner periodically obtains information on the {\em procFS} \cite{procfs-sysfs}  filesystem and records the peak stack usage of each thread in the \textit{Thread Information Table}, which will be discussed in the next section. Based on the recorded stack usage, the Stack Tuner automatically gives each thread suitable stack space to optimize the memory usage of the applications. For exact guidelines, this paper additionally configured three types of zones in the stack space: \textit{Low Zone}, \textit{Normal Zone}, and \textit{High Zone}\footnote{The default size of each zone is configured by the configuration file at the boot time of the CE/IoT device.}. Fig. \ref{fig:memory-tuner} shows how to classify the stack space of each thread. If the peak stack usage of a thread belongs to the \textit{Low Zone} or \textit{High Zone}, the Stack Tuner informs the developers, allowing them to fix the stack space requirement at the next compilation. To deliver this information, this paper modifies the Glibc \cite{glibc} library, which is well-established as the standard library for handling system calls in Linux. In the \textit{Low Zone}, the Stack Tuner points out that the thread is wasting the virtual memory space of the application. On the other hand, if the peak stack usage of the thread reaches the \textit{High Zone} range, the Stack Tuner generates the information that the thread may end up with a stack overflow in the near future. Finally, the Stack Tuner puts the \textit{Guard Page} at the end of the thread’s stack to detect a stack overflow.

\begin{figure}[ht]
\centering
\includegraphics[width=0.99\columnwidth,height=1.1in]{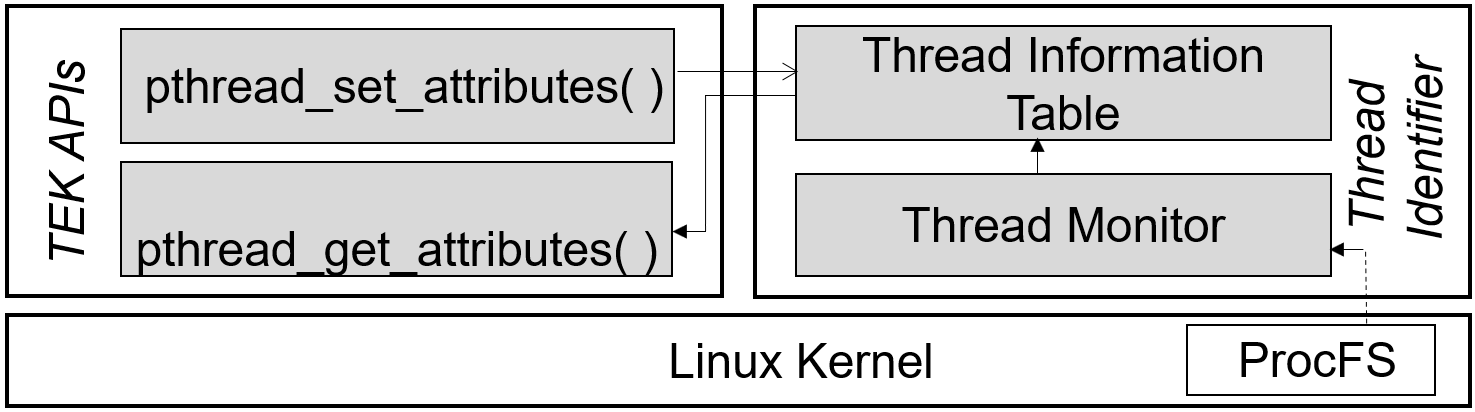}
\caption{The operation flow of the Enhanced Thread Identifier.}
\label{fig:thread-identifier}
\end{figure}

\subsection{Enhanced Thread Identifier and New APIs}\label{SS_thread_mm}
The main purpose of the Enhanced Thread Identifier is to easily identify the characteristics and attributes of each thread among hundreds of threads running on a CE/IoT device. In reality, an application calls the \textit{pthread\_create} API to create a new thread and the new thread just executes the thread function specified by the third argument of the \textit{pthread\_create} API \cite{rieker2006transparent, 2016rethink, engelschall2000portable, barney2009posix}. As a result, it is not easy to determine the role of the thread with the thread ID only. To enhance the identification of a thread, the Enhanced Thread Identifier records the information on a new thread created by an application into an auxiliary table, called \textit{Thread Information Table}.

Fig. \ref{fig:thread-identifier} shows the structure and relationship of the components of the Enhanced Thread Identifier. The Enhanced Thread Identifier extracts the thread attributes from the parameters of the \textit{pthread\_create} or \textit{pthread\_set\_attributes} API, and then records them in the \textit{Thread Information Table} on-the-fly. The thread attributes include the information on the role of the thread.

Application threads often call a function that connects to a sensor device to receive data from it (e.g., humidity sensor, temperature controller, air pressure sensor, or gas detection sensor). For example, a developer can set ``gas detection'' as the role of a thread with the \textit{pthread\_set\_attributes} API in order to easily lookup the gas detection thread running on a device. The \textit{Thread Monitor} in Fig. \ref{fig:thread-identifier} periodically collects thread information on the running threads (e.g., scheduling policy, scheduling priority, thread creation time, stack size, and virtual memory size) from the {\em ProcFS} \cite{procfs-sysfs} filesystem. The peak usage of the stack space of each thread is measured in this way. 

Meanwhile, this paper designed novel APIs to set or get the attributes of a thread in the \textit{Thread Information Table}. Application developers can mark a thread as time-critical or non-time-critical by triggering the \textit{pthread\_set\_attributes} API. When this function is called in the user space with a thread ID and its attributes, the Enhanced Thread Identifier searches for the thread ID in the \textit{Thread Information Table} and saves the thread attributes, including its priority and scheduling policy, into the table. On the other hand, the \textit{pthread\_get\_attributes} API is used to return the attributes of the thread.


\begin{table}
\caption{System configuration for experiments}
\begin{tabular}{c c c}
\hline
\hline
Content  & Item           & Specifications                      \\
\hline
\hline
H/W      & CPU            & Embedded CPU quad-core 1.2GHz   \\
         & RAM            & 1GB LPDDR2 (900MHz)                 \\
         & Storage        & 32GB MicroSD                      \\
         & Sensor Interface   &  GPIO-40 pin header           \\
\hline
S/W      & OS             & Linux 4.4.15 32bit (LTS)         \\
         & Virtual memory & 1 GB kernel space and 3 GB user space  \\
         & Compiler       & GCC 9.1                       \\
         & C library      & Glibc 2.29                      \\
         & Thread Model   & NPTL (Native POSIX Thread Library)~\cite{drepper2003native}  \\
\hline                                 
\end{tabular}
\label{table:testbed}
\end{table}

The \textit{Thread Information Table} requires additional memory space to store thread attributes. Considering that modern CE/IoT applications usually run more than 300 threads and the Enhanced Thread Identifier allocates 40 bytes to store the thread attributes for each thread, the Enhanced Thread Identifier requires just 12 KB (300 threads multiplied by 40 bytes) of additional memory space for 300 threads, and so, the additional memory cost is not significant. Also, considering that the read and write operations for managing the thread information are completed within 46 ns and 67 ns, respectively, when the Enhanced Thread Identifier saves the thread information into a typical memory device (1 GB LPDDR2 900MHz), it has little effect on the thread performance of the devices.

\section{Evaluation}\label{S_evaluation}

This section introduces an experimental environment and then explores how the proposed scheme, TEK, improves not only the response time of the time-critical threads but also the utilization of memory space. In particular, the evaluations in this section answer the following questions: (1) what is the difference between TEK and the conventional system in terms of context switching? (Section \ref{ss_ctx_switch}), (2) where does the improvement of the response time come from when TEK is enabled on CE/IoT devices? (Section \ref{ss_response_time}), and (3) how does TEK contribute to stack management at the kernel level? (Section \ref{ss_stack_mgt}).

\begin{figure}
\centering
\includegraphics[width=0.99\columnwidth,height=2.2in]{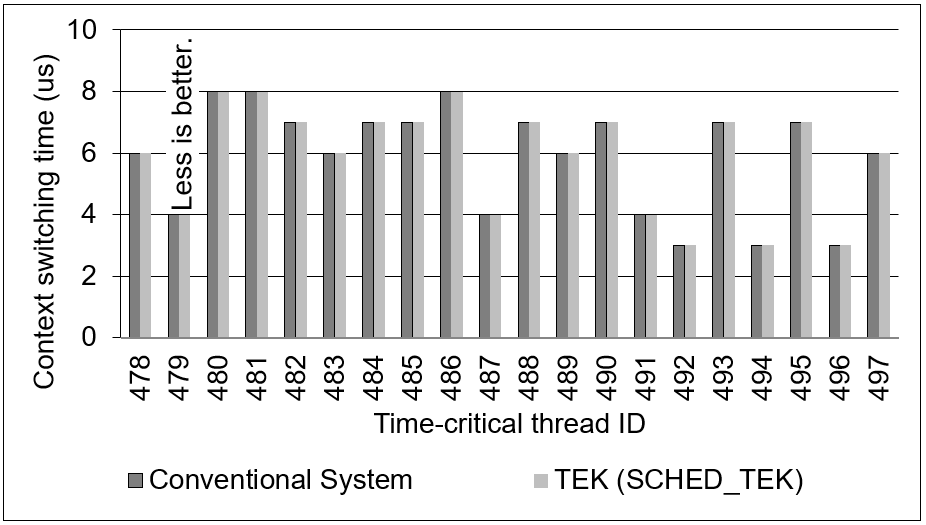}
\caption{The context switching time of the threads.}
\label{fig:eval-thread-ctx-cost}
\end{figure}

\subsection{Experimental Setup}

A prototype of TEK was implemented based on a commercial CE/IoT device using an Embedded CPU with 1 GB memory running Linux kernel 4.4. Table \ref{table:testbed} summarizes the evaluation setup in detail. The benefits of TEK were compared with the conventional kernel wherein the CPU scheduler provides coarse-grained control of threads and provides memory management based on a fixed-sized stack space. In this paper, all evaluations were conducted by categorizing threads as time-critical or non-time-critical so as to understand the performance difference of the events triggered by users. If a thread frequently handles user-level events during a short time period, it is considered to be time-critical because it requires a short response time. Otherwise, it is considered to be non-time-critical. The evaluation results were measured during the creation of 2000 threads after finishing the boot procedure.

\subsection{Context Switching of Threads}\label{ss_ctx_switch}

This paper first focuses on the performance of TEK in terms of context switching since a performance drop may be caused by the use of fine-grained thread management along with the thread’s priority. Fig. \ref{fig:eval-thread-ctx-cost} shows the experimental results for the context switching time. In the figure, the x-axis represents the thread ID of the time-critical threads, and the y-axis represents the context switching time of the time-critical threads. The proposed scheme has results similar to the conventional scheme even though it includes more behaviors for categorizing threads into time-critical and non-time-critical ones. This is possible because the proposed scheme offloads the operations for thread classification to the CFS scheduler by logically migrating the threads between the regions implemented with the doubly-linked list as depicted in Section \ref{SS_cpu_mediator}.

\begin{figure}
\centering
\includegraphics[width=0.99\columnwidth,height=1.8in]{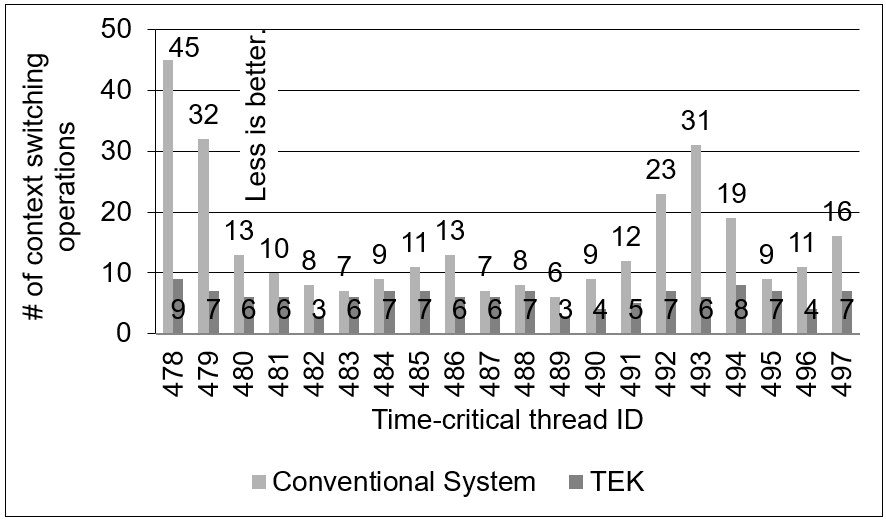}
\caption{The frequency of context switching operations on time-critical threads.}
\label{fig:eval-thread-ctx-freq}
\end{figure}

\begin{figure}
\centering
\includegraphics[width=0.99\columnwidth,height=2.0in]{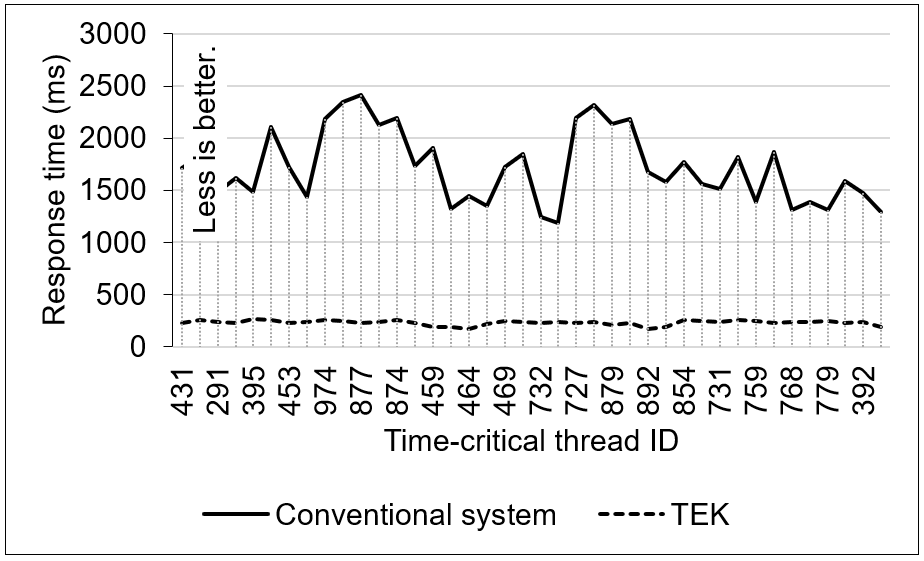}
\caption{The evaluation results of the user-space SCHED\_TEK policy for improving user responsiveness of time-critical threads under CPU contention.}
\label{fig:eval-time-critical-thread-response-time}
\end{figure}

\begin{figure*}
\centering
\includegraphics[width=1.9\columnwidth,height=2.0in]{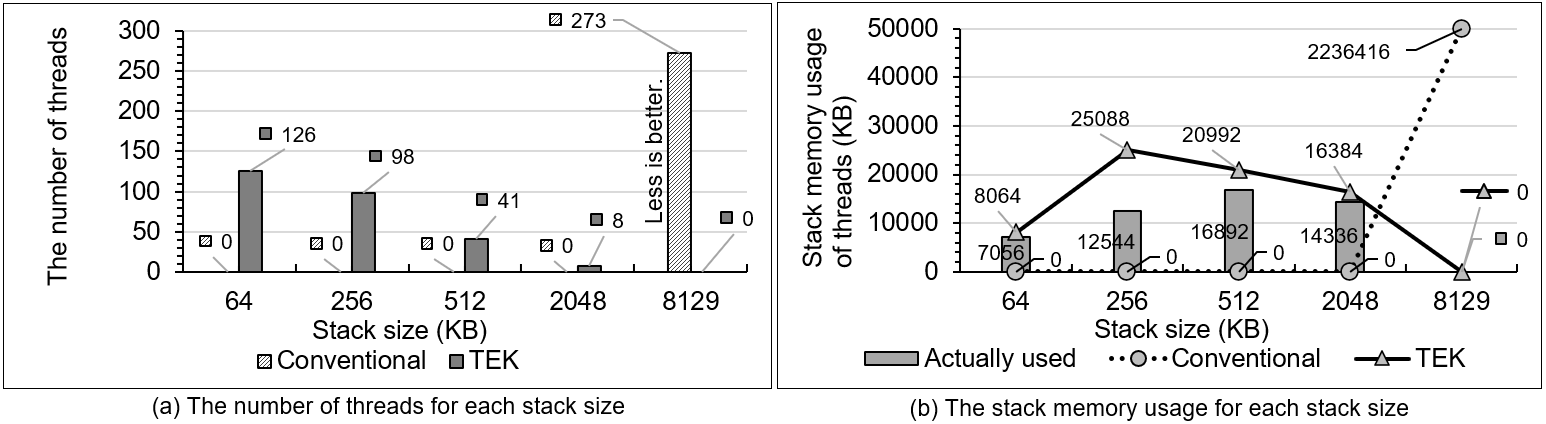}
\caption{Virtual memory consumption for each stack size of the threads}
\label{fig:eval-stack-size}
\end{figure*}

\begin{figure*}
\centering
\includegraphics[width=1.9\columnwidth,height=1.8in]{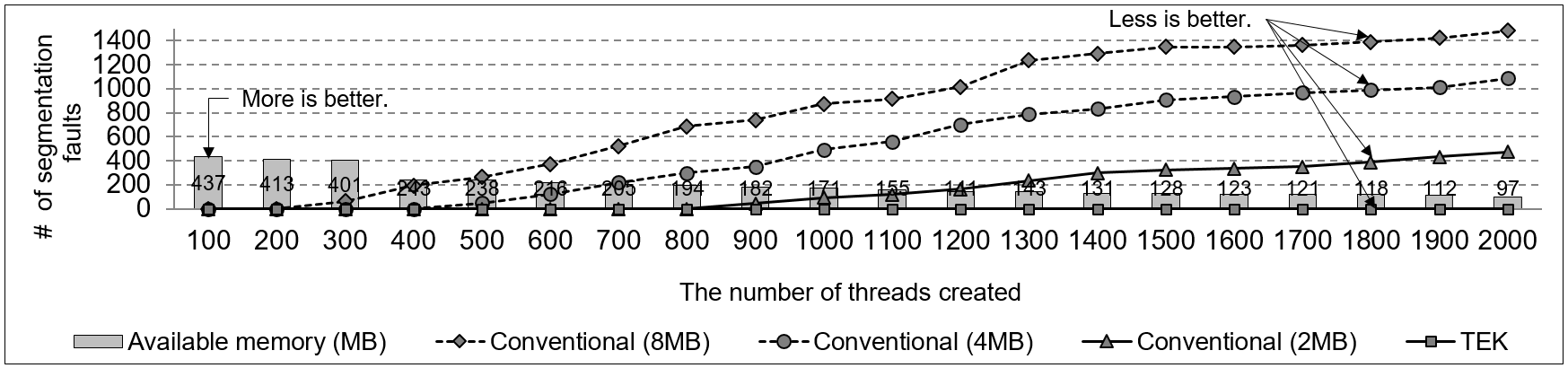}
\caption{The experimental results for segmentation fault frequency while creating threads.}
\label{fig:eval-thread-segfaults}
\end{figure*} 

Meanwhile, Fig. \ref{fig:eval-thread-ctx-freq} shows the number of context switching operations centered around the time-critical threads. This figure confirms that TEK significantly reduces the number of context switching operations compared with the conventional method. In the best case, TEK reduces the number of context switching operations by up to 41\%. The reason behind this is that TEK assigns more CPU time to the time-critical threads by isolating them from the group scheduling policy. As a result, the time-critical threads can speed up their response time by up to 42\% compared with those of the conventional system. In summary, TEK provides more opportunities to time-critical threads in terms of CPU scheduling with little overhead. In addition, TEK does not require any modifications to the conventional one-to-one thread model because it uses the API of the conventional thread model.

\subsection{Response Time of User-Level Event}\label{ss_response_time}

This section discusses the improvement obtained to the response time of the time-critical threads using the TEK user space thread manager. To measure the response time of the time-critical threads, the experiment was conducted under a real-life scenario that included two steps. As described in Section \ref{SS_cpu_mediator}, modern operating systems can limit the CPU usage of the running threads by classifying them into four different scheduling groups: urgent, normal, service, and background \cite{bellasi2015cgroup}. Now, the first step in the scenario creates threads that are evenly deployed to each thread group so as to create a CPU intensive situation. Then, the second step measures the response time of each thread when it handles a user-level event, such as a touch screen input on the CE/IoT device. In other words, the response time of each time-critical thread was measured during 100\% CPU utilization.

Fig. \ref{fig:eval-time-critical-thread-response-time} shows the evaluation results of TEK compared with the conventional system. As shown in the figure, the conventional system leads to latency fluctuations; the average response time was 1719 ms when all of the time-critical threads competed for the CPU. This is because the non-time-critical threads can frequently stagnate and even hang on the time-critical threads. Alternatively, TEK shows steady performance results, dramatically reducing the average response time of the time-critical threads by up to 235 ms. These results are meaningful because there was a lot of competition for CPU resources in the thread group. The reason behind such significant improvements is that the CPU Mediator efficiently isolates and handles time-critical threads in terms of the CPU resources. Unfortunately, the CPU Mediator causes a negative impact on the performance of non-time-critical threads because they can be preempted to yield resources to the time-critical threads. In the worst case, the response time of the non-time-critical threads was stalled by up to 1487 ms. However, it is important to mention that the non-time-critical threads, such as software update threads, reserved task threads, and system management threads, do not react to user activities on-the-fly.

\subsection{Stack Management of Threads}\label{ss_stack_mgt}

Generally, whenever one thread is created, the memory manager in the kernel allocates a fixed-size chunk of stack memory for the created thread. Unfortunately, if the thread uses a smaller region of memory than the allocated chunk, the unused memory space is wasted. This unused memory space may indirectly cause a segmentation fault because it leads to a shortage of free memory. On the other hand, as mentioned, TEK efficiently allocates a stack memory chunk that best fits the thread using the Stack Tuner. Fig. \ref{fig:eval-stack-size} shows the accumulated usage of stack memory allocated to the created thread. For accurate evaluation, the data in Fig. \ref{fig:eval-stack-size} were monitored during creation of a total of 273 threads over 15 days. As expected, Fig. \ref{fig:eval-stack-size}--(a) clearly confirms that TEK uses a much smaller amount of memory space than the conventional system. Also, TEK allocated only 70 MB of memory, even though a total of 273 threads were running simultaneously, as shown in Fig. \ref{fig:eval-stack-size}--(b).

Fig. \ref{fig:eval-thread-segfaults} plots the number of segmentation faults that actually occurred while running the threads on the experimental CE/IoT device. As mentioned before, the conventional systems employ a fixed-sized chunk of memory to support the creation of a new thread, and therefore, the evaluation was performed by varying the size of the fixed-sized stack space from 2 MB to 8 MB. In the conventional system, the evaluation results clearly show that the number of segmentation faults increased significantly with an increase in the number of threads. In particular, after the number of accumulated threads reached 200, the conventional system became unstable and could not guarantee a stable response for the creation of a new thread because of the exception handling of the segmentation fault. On the other hand, TEK maintained good conditions over most of the experiment because the proposed system supports a fine-grained stack allocation that exactly allocates stack memory space for the amount actually used in the thread. Of course, TEK also wastes a small portion of memory because of page alignments. To understand how much memory space is wasted, the amount of allocated memory usage was monitored on both the TEK and the conventional system. As shown in Fig. \ref{fig:eval-stack-size}, TEK only used 39\% (70528 KB) of the stack memory space thanks to the Stack Tuner, while the conventional system used 4300\% (2236416 KB), compared with the stack memory space actually used by the threads (50828 KB).

\section{Related Work}\label{S_related_work}
This section summarizes prior work to clearly understand the difference between the proposed scheme and the conventional system in terms of the thread model, thread performance, and thread management.

\textit{Thread Model and Performance:} Many studies have been performed to enhance the thread model. NPTL \cite{drepper2003native} pointed out the issue of the scalability of the Linux scheduler, then proposed the O(1) Linux scheduler to address the issue both on multi-core and single-core architecture. In particular, NPTL designed a FUTEX synchronization mechanism to support the one-to-one thread model without additional overhead. Wong \cite{wong2008fairness} presented the CFS scheduler to ensure a fair allocation of CPU resources to tasks without sacrificing interactive performance. As a result, this scheduler could replace the O(1) scheduler \cite{wong2008fairness} in the Linux kernel. Meanwhile, PK \cite{miller1999pk} focused on the performance of threads and proposed a concurrency model based on POSIX threads (Pthreads) to improve thread performance, including real-time threads. Engelschall \cite{engelschall2000portable} described a portable multi-threading mechanism that supports the expeditious creation and execution of threads during the simultaneous execution of multiple threads. In addition, to achieve backward compatibility, Engelschall developed a mechanism based on ANSI-C on the Unix system. In summary, all of the above studies focused on improving the performance of threads in high-performance computing environments equipped with large-scale hardware resources. Therefore, they are different from TEK in that TEK considers small-scale hardware environments, like CE/IoT devices.

\textit{Thread Management:} Adya \cite{adya2002cooperative} focuses on cooperative task management to guide the concurrency conditions of the system for program architects. A prototype of the cooperative task management method was implemented based on the event-driven approach so as to meet the requirements of thread concurrency. On the other hand, Arachne \cite{qin2018arachne} addresses low-latency and high-throughput applications by designing short-lived threads. In this scheme, threads running in the user level are handled with core-aware scheduling, which assigns the cores to each thread according to the application requirements. In other words, the desired scheduling can be achieved with core-aware thread management. However, since the APIs are not POSIX compatible, it is difficult to immediately port them to modern CE/IoT devices.

\section{Conclusion}\label{S_conclusion}

Contemporary CE devices, which have sensor and network modules, have unique characteristics compared with general desktop or server systems in that the threads of an application must be handled by limited resources, such as low clock speed CPU and small capacity memory. This allows for low power requirements, miniaturization, and cost competitiveness. This paper targeted enhancing the existing one-to-one thread model to resolve the resource management problems of the user-space threads on low-end CE/IoT devices. To handle threads more efficiently in these systems, this paper proposed state-of-the-art resource management facilities for CE devices: a CPU Mediator, Stack Tuner, and Enhanced Thread Identifier. This paper shows that the proposed system dramatically improves the response time of time-critical threads by up to 7x and saves available virtual memory space by up to 3.4x. In addition, the proposed system supports a POSIX-compatible thread scheduling API that allows developers to run unified applications on both small-scale and large-scale hardware platforms. Also, the proposed system supports a light-weight system resource manager to improve naive stack management on the low-end CE devices.



\bibliographystyle{IEEEtran} 


\bibliography{ref}



\vspace{-3em}
\begin{IEEEbiography}[{\includegraphics[width=1in,height=1.25in,clip,keepaspectratio]{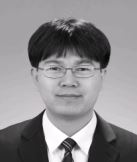}}]{Geunsik Lim}
received his B.S. degree in Computer Science and Engineering from Ajou University, in South Korea in 2003. He received his M.S. degree in the College of Information and Communication Engineering from Sungkyunkwan University, in South Korea in 2014. He is currently a Ph.D. student in the Department of Electrical and Computer Engineering, Sungkyunkwan University, and also a principal software engineer for Samsung Electronics in South Korea. His current research interests include system optimization, operating systems, software platforms, and on-device artificial intelligence.
\end{IEEEbiography}
\vspace{-3em}
\begin{IEEEbiography}[{\includegraphics[width=1in,height=1.25in,clip,keepaspectratio]{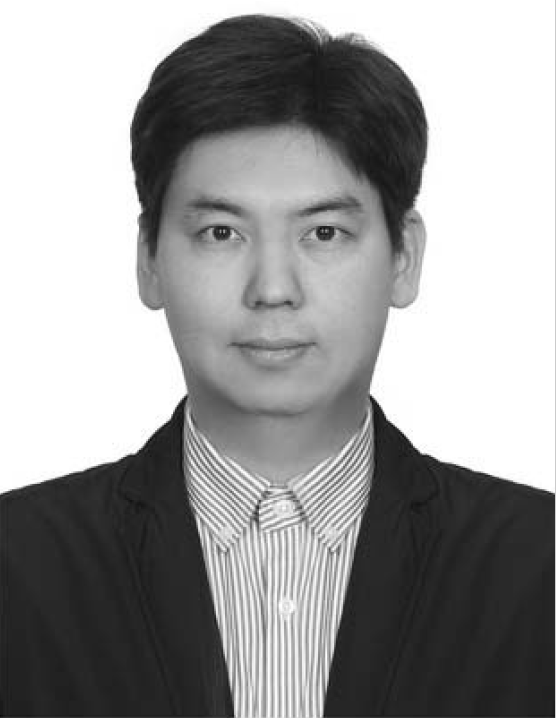}}]{Donghyun Kang}
is an assistant professor with the Department of Computer Engineering at Changwon National University in South Korea. Before joining Changwon National University, he was an assistant professor with Dongguk University (2019-2020) and a software engineer at Samsung Electronics in South Korea (2018-2019). He received his Ph.D. degree in College of Information and Communication Engineering from Sungkyunkwan University in 2018. His research interests include file and storage systems, operating systems, and emerging storage technologies.
\end{IEEEbiography}
\vspace{-3em}
\begin{IEEEbiography}[{\includegraphics[width=1in,height=1.25in,clip,keepaspectratio]{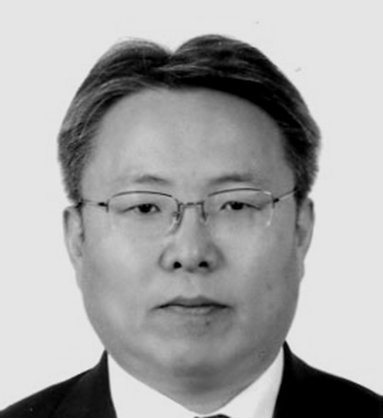}}]{Young Ik Eom}
received his B.S., M.S., and Ph.D. degrees in Computer Science and Statistics from Seoul National University, South Korea, in 1983, 1985, and 1991, respectively. Since 1993, he has been a Professor with Sungkyunkwan University, South Korea. From 2000 to 2001, he was a visiting scholar with the Department of Information and Computer Science, University of California at Irvine. He also was president of the Korean Institute of Information Scientists and Engineers in 2018. His research interests include virtualization, operating systems, file and storage systems, cloud systems, and UI/UX system.
\end{IEEEbiography}

\end{document}